\documentstyle[prl,aps,twocolumn]{revtex}
\input epsf
\newcommand{\damb}{{d+2}} \newcommand{\da}{{d+1}} 
\newcommand{\dc}{{d}} \newcommand{\dm}{{d-1}}
\newcommand{\twod}{1} 
\newcommand{\adsd}{\mbox{{AdS}$_{1,\dc}$}}  
\newcommand{\RR}{{\mathbf{R}}}
\newcommand{\sod}{\mbox{SO$(2,\dc)$}}
\begin{document}
\draft 
\title{Local Quantum Observables in the Anti-deSitter - Conformal QFT
  Correspondence} 
\author{Karl-Henning Rehren}  
\address{Institut f\"ur Theoretische Physik,
Universit\"at G\"ottingen, 37073 G\"ottingen, Germany}
\date{March 15, 2000}
\maketitle
%
%
\begin{abstract} 
Quantum field theory on $\da$-dimensional anti-deSitter space-time admits a
re-interpretation as a quantum field theory with conformal symmetry on
$\dc$-dimensional Minkowski space-time. This conjecture originally emerged
from string theory considerations. Here, it is proven in a general
framework by an explicit identification between the local observables of
the two corresponding theories.   
\end{abstract}
\narrowtext 
\section{The AdS-CFT correspondence}
Quantum field theory on anti-deSitter space-time (AdS) has received an
important impact from string theory. Evidence was found \cite{M,W,GK}
to the effect that certain theories on $\da$-dimensional AdS
equivalently describe conformally invariant quantum field theories
(CFT) on $\dc$-dimensional Minkowski space-time. In particular, the 
higher-dimensional AdS theory can be recovered from the
lower-dimensional CFT. So the correspondence is ``holographic'' in the
sense of \cite{tH} where the influence of a black hole horizon on
quantum fields in the ambient bulk space was discussed.  

This correspondence has attracted much attention,
as it suggests a wealth of implications for quantum gravity and for
gauge theories in physical (four-dimensional) space-time. For a
comprehensive review, see \cite{AMO}. 

A large portion of the work on the AdS-CFT correspondence
crucially involves ``stringy'' pictures (branes, duality, M-theory) when
comparing contributions to the relevant path integrals and/or
correlation functions. The correspondence is, however, claimed to be a
model-independent feature of quantum field theory \cite{W}. 

So the question arises as to whether it can be understood in more
basic terms not relying on string theory. In fact, one main
ingredient, the coincidence of the space-time symmetry groups, is even
a purely classical one, long known to physicists: both the isometry
group of AdS in $\da$ dimensions and the conformal group of Minkowski
space-time in $\dc$ dimensions are \sod.   

It is the aim of this letter to show that it is indeed possible to
understand (and to prove) the AdS-CFT correspondence in a general
quantum field theoretical set-up. We shall first give a brief
introduction to this set-up in Sect.\ II. It is certainly the most
general one to incorporate the two fundamental principles of
relativistic Covariance and Causality in quantum theory.   

\section{Local observables}

The prime objects of consideration in a quantum theory are the quantum  
observables, represented as self-adjoint operators on a Hilbert space
whose elements are the vector states in which the system can be
prepared. The real expectation values of the observables in various 
states (e.g., the vacuum state) predict the statistical outcome of any
measurement. 

In a relativistic quantum theory, in contrast to quantum mechanics, 
observables have the property of {\em localization}, compatible with 
Locality and Covariance. Locality is a consequence of Einstein Causality
and means that observables which are localized at spacelike distance
commute with each other. Covariance requires that the space-time
symmetry group acts (by unitary operators $U(g)$)
on the localization of observables according to
its geometric significance, thereby preserving any algebraic relations
among them. In the AdS-CFT case at hand this group is the AdS resp.\
conformal group \sod.  

In conventional quantum field theory, the above features are usually
encoded by quantum fields: objects $\phi(x)$ localized at the points
$x$ in space-time, commuting at spacelike distance and transforming
under some relativistically covariant transformation law. Due to the
singular nature of quantum fields, these are not Hilbert space
operators, but become operators after smearing with a test
function. The best localization of observables is therefore an open
region in space-time which contains the support of a test function.  

The choice of quantum fields used for the description of a
relativistic quantum system is to a large extent a matter of
convenience. It has been recognized long ago \cite{Bo} that different
quantum fields may well describe the same quantum system. Prime
examples are the equivalence of the Sine-Gordon and Thirring model
\cite{Col}, the re-interpretation of Chern-Simons theories in terms of
models with Yukawa interaction \cite{L}, and the duality of certain
supersymmetric Yang-Mills theories \cite{MO,SW}. 

One is thus led to the conclusion \cite{HK} that what determines the
physical interpretation of a quantum theory are not the individual quantum
fields but the algebras of Hilbert space operators which are generated
by localized field operators. Theories with possibly different
equations of motion may well be equivalent if they only generate the
same system of local algebras. The existence of generating fields is
not even required if the local algebras can be specified by any other
consistent prescription.   

To conclude, the knowledge of localization is sufficient for the physical
interpretation of a theory. The more specific interpretation of
individual observables can be recovered from their correlations with
other localized observables (exhibited in expectation values). This
insight has proved most fruitful for a wide spectrum of structural
results, ranging from scattering theory, a clarification of the
superselection (charge) structure, to an algebraic renormalization
group analysis. For a recent review, see \cite{BH}.      

As we are aiming at a general and intrinsic description of the AdS
resp.\ CFT theories, we shall deal here with their local algebras and
assume that they comply with the requirements of Covariance and
Locality, as well as the additional but obvious property of Isotony:
an observable localized in some region $O$ is localized in any larger
region also. 

\section{AdS-CFT resumed}
We adopt the set-up, sketched above, of algebras of localized
observables, based on fundamental principles generally accepted. It
applies to any physically reasonable relativistic quantum field
theory, including certain string theories \cite{D}. We shall show that
it is the most natural set-up to establish the AdS-CFT correspondence. 
For it is this structure which is preserved by the correspondence. In
contrast, the description in terms of specific fields and Lagrangeans
will in general not be preserved.   

As explained, a quantum field theory is specified if the algebras $A(O)$ 
of observables localized in each open space-time region $O$ are known:  
$$A(O)={\rm span}\{\hbox{$\phi$: $\phi$ is an observable localized in $O$}\}.$$
This assignment has to comply with Covariance and Locality. 

Thus, to prove the AdS-CFT correspondence, we have to establish a
prescription specifying the algebras $B(W)$ of local AdS observables
for suitable AdS regions $W$ if the algebras $A(K)$ of local CFT
observables for suitable Minkowski regions $K$ are given, and {\em vice
versa}. This prescription must pass on Locality and Covariance from the
given theory to the new theory in correspondence.  

As the discussion of quantum fields in the presence of a gravitational
horizon \cite{tH,M} underlying the holographic picture suggests, the
set of all operators representing observables should be the same for both
theories, and act on the same Hilbert space. Moreover, the conformal
space-time should play the role of a horizon in AdS space-time.  

Indeed, the $\da$-dimensional AdS space-time given as
$$\adsd=\{\xi\in\RR^\damb : 
\xi_0^2 - \xi_1^2 - \dots - \xi_\dc^2 + \xi_\da^2 = R^2 \}$$ 
has a ``boundary'' at spacelike infinity, and the induced (properly
rescaled) metric of the boundary is that of $\dc$-dimensional
conformal Minkowski space-time. The action of the AdS group on \adsd\
preserves the boundary, and acts on it like the conformal group in
$\dc$-dimensional Minkowski space-time.    

The law of causal propagation between the bulk of AdS and its
boundary suggest how to find the prescription to identify localized
observables between the two theories \cite{KHR1}. Namely, let $K$ be a
causally complete open and convex region in 
Minkowski space-time, -- a convenient choice is a double-cone, i.e.,
the intersections of a future-directed and a past-directed
light-cone. It uniquely determines a wedge-shaped region $W$ of  
AdS which consists of all points at which one can receive signals from
some point of $K$, and from which one can send signals to some other
point of $K$ (the ``causal completion'' of $K$ in AdS). Conversely,
the boundary region $K$ is recovered from this AdS region $W$ by
taking its intersection with the boundary of AdS.

We omit proofs of the geometric facts mentioned here and in the
sequel; the reader may find details in \cite{KHR1}. It is largely sufficient 
to visualize \adsd\ in suitable coordinates as a full cylinder 
$\RR\times B^{\dc}$ whose axis $\RR$ represents time (possibly 
periodic, see below), and whose boundary $\RR\times S^{\dm}$
represents spacelike infinity. ($B^\dc$ is a ball, and $S^\dm$ is a
sphere.) Double-cones $K$ are inscribed into the boundary, and the
wedges $W$ look like actual wedges ``chopped into the cylinder'' (cf.\
Fig.\ 1).
\begin{figure}[hbt]
\epsfxsize81mm
\epsfbox{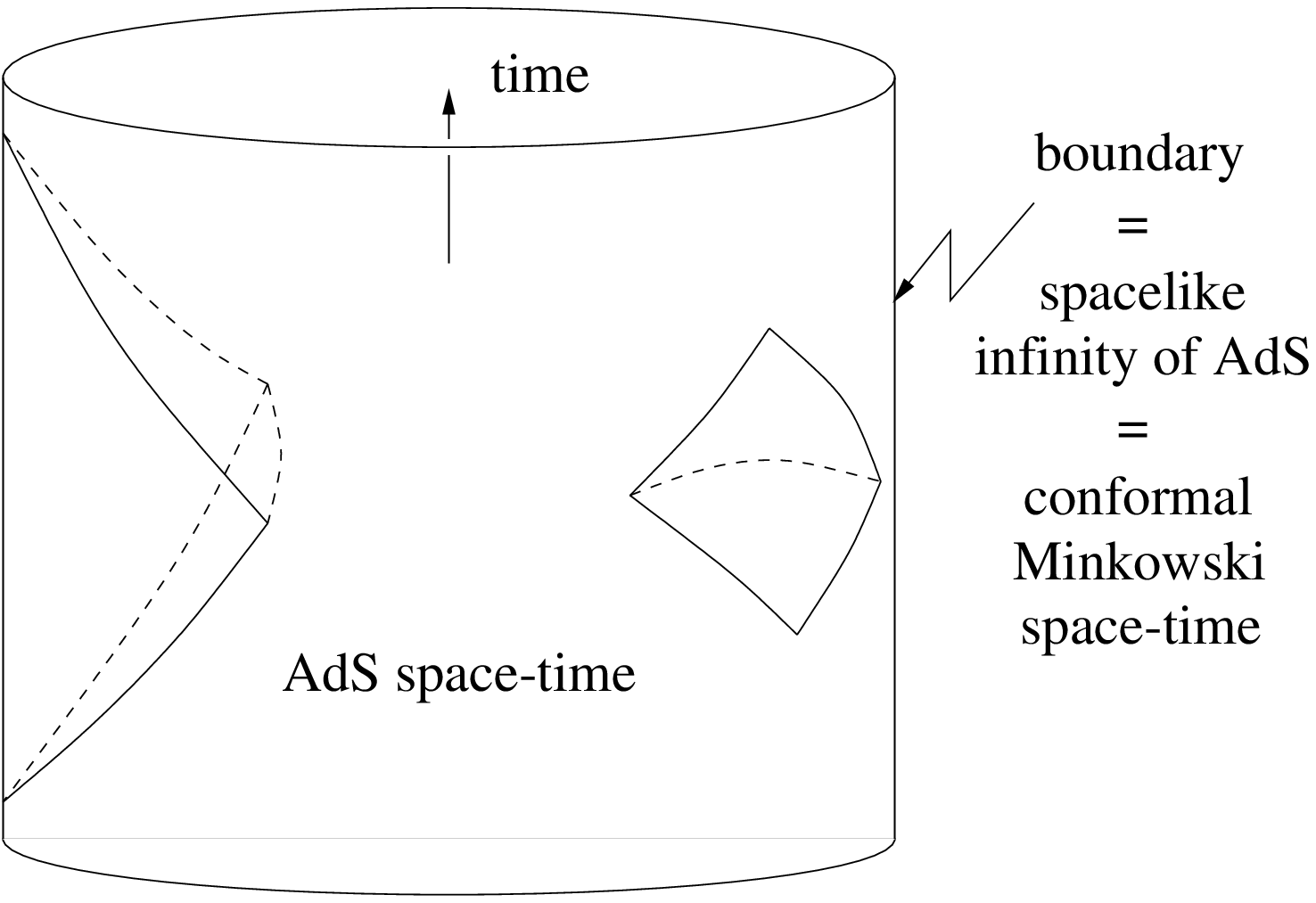}
\end{figure}
FIG.\ 1. Wedge regions in AdS and corresponding double-cones in
  the boundary (in Penrose coordinates) \vskip2mm

Let us denote the bijective correspondence between AdS wedges and boundary 
(Minkowski) double-cones by $K=\iota(W) \Leftrightarrow W=\iota^{-1}(K)$. 
Then the specification  
$$A(K):=B(W) \qquad {\rm if} \qquad K=\iota(W),$$ 
determines a system of local algebras $A(K)$ of observables on
Minkowski space-time, given the system of local algebras $B(W)$ on AdS. 

This identification preserves Covariance. For if a wedge $W$ is transformed
under the AdS group, then its intersection $K$ with the boundary
undergoes a conformal transformation, and {\em vice versa}. More concretely,
if $g$ stands for an element of the AdS group, and $\dot g$ for its
induced conformal transformation of the boundary, then
$$\iota(gW)=\dot gK \quad  {\rm if}\quad K=\iota(W), $$ 
implying the correct transformation of the observables:
\begin{eqnarray*} 
 U(g)A(K)U(g)^{-1}=U(g)B(W)U(g)^{-1}= \\ =B(gW)=A(\iota(gW))=A(\dot gK).
\end{eqnarray*}
In particular, the conformal symmetry is implemented by the same
unitary Hilbert space representation $U$ of \sod\ as the AdS symmetry.

The identification also preserves Isotony: One has 
$$\iota(W_1)\subset \iota(W_2)\qquad {\rm if} \qquad W_1 \subset W_2$$
for obvious reasons. Since the given algebras $B(W)$ satisfy Isotony,
this implies
$$A(K_1)\subset A(K_2) \qquad {\rm if} \qquad K_1 \subset K_2.$$

Finally, also Locality is preserved: Namely, $\iota$ maps pairs of causally
complementary AdS regions onto pairs of causally complementary
boundary regions (the causal complement $X'$ of a region $X$ consists
of all points which are spacelike separated from any point in $X$): 
$$\iota(W')=K'  \qquad {\rm if} \qquad \iota(W)=K.$$ 
If now $K=\iota(W)$ and $\hat K$ are spacelike separated, then 
$\hat K$ is a subset of $K'=\iota(W')$. Observables localized in $K$
and $\hat K$ are thus identified with operators in $B(W)$ and $B(W')$, and 
hence commute as required by Locality.         

Since $\iota$ is a bijection, the prescription can be reversed, specifying 
the system of algebras $B(W)$ by the given local algebras $A(K)$:   
$$B(W):=A(K) \qquad {\rm if} \qquad W=\iota^{-1}(K).$$ 
By the same arguments as before, Covariance, Isotony and Locality
hold for $B(W)$ if they hold for $A(K)$.

We emphasize that we have reduced the problem of establishing the AdS-CFT
correspondence to a completely geometric one. It is not necessary to
proceed from a specific quantum field theory in order to understand
why it admits a holographic re-interpretation.  

We now turn to discuss some physical implications of the
correspondence thus established.

\subsection{Change of the physical interpretation}

Although the pair of corresponding theories shares the same set of local
observables as operators on the same Hilbert space, they have different 
physical interpretations. This possibility is familiar from quantum
mechanics where the state space is always a separable Hilbert space
and the set of all observables are the functions of position and momentum. 

The physical interpretation arises from the assignment of observables
to localization regions, and the consequent correlations in the
expectation values of observables in various geometric arrangements.
A re-assignment completely changes the interpretation. For instance, 
the description of a scattering experiment would require the
determination of correlations between observables at asymptotically
large distances. As notions like ``spacelike infinity'' are not
preserved by the bijection $\iota$, the computation of asymptotic
correlations yields entirely different results in corresponding theories.  

Furthermore, the one-parameter subgroups of \sod\ describing time
translations in the AdS and conformal interpretations do not
coincide. Therefore, also notions like dynamics, energy and entropy
change their meaning under the AdS-CFT correspondence.   

\subsection{Pointlike AdS and extended CFT observables}

Not even the concept of a point is preserved by the correspondence
(which should not be a surprise since corresponding theories live in
space-times of different dimension). For instance, arbitrarily small 
double-cones in the boundary correspond to wedges close to infinity in
AdS, which always have infinite volume. That an observable can be
written as a field $\phi(x)$ smeared with a test function, or as some
function of field operators, may be true in AdS, but not in the
corresponding CFT, or {\em vice versa}. Thus, a description in terms of
fields may fail in one of the two theories. This is an instance where
the advantage of thinking in terms of extended observables and the
description of their time evolution by an automorphism group in
contrast to fields and equations of motion, is clearly exhibited. 

We want to demonstrate that the identification of localized
observables implies that AdS observables localized in finite AdS
regions (in particular proper AdS {\em fields}) correspond to
genuinely extended CFT observables. In the argument we assume the
dimension of AdS to be $\da>1+1$ (the case $d=\twod$ is very special
and has been discussed elsewhere \cite{KHR1,KHR2}).  

Let $X$ be a bounded region in AdS. Pick some wedge $W$ which
contains $X$ and consider the family of wedges $W_i$ contained in $W$
which are spacelike to $X$. One finds that the
corresponding boundary double-cones $K_i=\iota(W_i)$ are contained in
$K=\iota(W)$ and cover its $t=0$ surface.
 
Let $B(X)$ denote the algebra of AdS observables localized in 
$X$. It belongs to $B(W)$ and commutes with all observables in
$B(W_i)$, hence as a CFT observable it belongs to $A(K)$
and commutes with all observables in $A(K_i)$. It commutes in
particular with all boundary fields smeared over a neighborhood of a
Cauchy surface of $K$.      

Now, if the algebra $A(K)$ were generated by the family of subalgebras
$A(K_i)$, we would find that $B(X)$ belongs to $A(K)$ and commutes
with every operator in $A(K)$, and therefore is a commutative
algebra. Its elements can be classical observables only. This
cunclusion applies, e.g., if the CFT is completely described by fields
with a causal dynamical law, since the fields along the Cauchy surface
generate all observables localized in $K$.

Reversing the argument, we conclude that the {\em quantum} observables in
$B(X)$ (e.g., AdS field operators smeared within $X$) correspond to CFT
observables in $A(K)$ which are {\em not} generated by the family of
subalgebras $A(K_i)$ covering the $t=0$ surface of the double-cone
$K$. They are thus genuinely extended CFT observables. In particular,
the CFT cannot be {\em completely} described by its fields with a
dynamical law.        

The extended observables of the CFT, whose presence is implied by the
above argument, might be Wilson loop operators in nonabelian gauge
theories. While observable fields fail to generate all quantum
observables of the CFT, gauge invariant nonlocal ``functions'' of
gauge fields could account for the rest. 

On the other hand, CFT fields correspond
to AdS observables attached to infinity, which might just be suitably
renormalized limits of AdS fields \cite{BBMS}. More enthralling is the
possibility to identify some AdS degrees of freedom, which
collectively restore the crossing symmetry of conformal operator
product expansions obtained by an AdS prescription \cite{HPR}, as 
{\em strings} -- thus making contact with the original conjectures
\cite{M,W,GK}.      

\subsection{Global structure of space-time}
The bijection $\iota$ between double-cones and wedges (Sect.\ III)
pertains to proper conformal Minkowski space-time and projective AdS
space-time which is the AdS hyperboloid with antipodal points $\xi$
and $-\xi$ identified: $\rm P\adsd=\adsd/(\xi\sim-\xi)$. One may still
formulate the AdS theory on \adsd, but then one
finds $B(W)=B(-W)$: antipodal wedges have the same observables.  

A CFT on proper conformal Minkowski space-time cannot describe any
interaction since its observables commute also at timelike separation,
hence any causal influence is bound to lightlike {\em geodesic}
propagation. This implies that observables in the corresponding theory
on projective AdS commute unless their localizations are connected by
a causal {\em geodesic} (see also \cite{BFS} for an independent
argument to the same effect). But if causal influence only propagates
along geodesics, then no process like the decay of a particle (with
non-geodesic trajectory due to recoil) is possible. Hence, the AdS
theory will also not describe a system with interaction.     

Theories of physical interest thus rather ``live'' on covering spaces
of projective AdS and of conformal Minkowski space-time, respectively,
where the closed timelike curves of these manifolds are unwinded. 
The bijection $\iota$ generalizes to {\em corresponding} 
covering spaces, and especially the universal coverings of both spaces
(disregarding the case $d=\twod$ which is again peculiar \cite{KHR1}).   

This allows for possible anomalous dimensions of conformal fields
and nontrivial timelike commutation relations, and evades the above
conclusion of geodesic propagation and absence of interaction on AdS. 

\section{Conclusion}
We have obtained the proper identification of local quantum
observables which underlies the ``holographic'' correspondence between 
quantum field theory on $\da$-dimensional anti-deSitter space-time and 
$\dc$-dimensional conformal quantum field theory. It simply reflects
the geometric law of causal propagation between AdS space-time and its
boundary. But it suffices to define one theory in terms of the other,
and entails a complete re-interpretation of the physical content.  

We conclude from this result, among other things, that AdS {\em fields}
correspond to genuinely {\em extended} CFT observables. These can
be a hint at conformal gauge theories.  

Strings play no particular role in the present explanation of the
AdS-CFT correspondence. But it is conceivable that they
re-appear as ``collective'' AdS variables required by crossing
symmetry of the corresponding CFT.

\end{document}